\documentclass[prl,aps,twocolumn,superscriptaddress,showpacs,preprintnumbers,amssymb]{revtex4}

\usepackage{color}
\usepackage{graphicx}% Include figure files
\usepackage{ulem}
\usepackage{float}
\usepackage{here}
\usepackage{bm}
\usepackage{amstext}
\usepackage{amsfonts}
\usepackage{dcolumn} % Align table columns on decimal point
\usepackage[colorlinks=true,urlcolor={blue},citecolor={blue}]{hyperref}
\usepackage{pgf}
\usepackage{graphicx}
\usepackage[T1]{fontenc}

\def\textbf#1{\boldsymbol{#1}}

\def\i{\mathrm{i}\,}

\begin{document}

\title{Soft phonon driven orbital order in CaMn$_{7}$O$_{12}$}
\author{S. M. Souliou}
\affiliation{European Synchrotron Radiation Facility, BP 220, F-38043 Grenoble Cedex, France}
\author{Y. Li}
\affiliation{International Center for Quantum Materials, School of Physics, Peking University, Beijing 100871, China}
\affiliation{Collaborative Innovation Center of Quantum Matter, Beijing 100871, China}
\author{X. Du}
\affiliation{International Center for Quantum Materials, School of Physics, Peking University, Beijing 100871, China}
\author{M. Le Tacon}
\affiliation{Karlsruhe Institute of Technology, Institut f\"{u}r Festk\"{o}rperphysik, D-76021 Karlsruhe, Deutschland}
\author{A. Bosak}
\affiliation{European Synchrotron Radiation Facility, BP 220, F-38043 Grenoble Cedex, France}

\date{\today}

\begin{abstract}
We use variable-temperature x-ray thermal diffuse scattering and inelastic scattering to investigate the lattice dynamics in single crystals of multiferroic CaMn$_{7}$O$_{12}$ which undergo a series of orbital and magnetic transitions at low temperatures. 
Upon approaching the charge and orbital ordering temperature T$_o$=250 K from above, we observe intense diffuse scattering features and a pronounced optical phonon softening centered around the superstructure reflections of the incommensurately modulated structure.
The phonon anomaly appears well above T$_o$ and continuously increases upon cooling, following a canonical power-law temperature dependence that confirms the transition at T$_o$ to be of second order and related to a soft-phonon lattice instability.
Microscopic mechanisms for the incommensurate charge and orbital ordering based on competing interactions and on momentum-dependent electron-phonon coupling could both account for the observed extended momentum width of the phonon softening. 
Our results highlight the importance of the lattice interactions in the physics of this magnetically induced ferroelectric system. 

\end{abstract}


%%%%%%%%%%%%%%%%%%%%%%%%%%%%Introduction%%%%%%%%%%%%%%%%%%%%%%%%%%%%%%%%%%%%%%%%%%%%%%%%%%%%%%%%
\maketitle
\section{\label{sec:Introduction}I. Introduction}

Mixed-valence manganese oxides offer an ideal platform for the study of intertwined orbital, magnetic, charge and lattice degrees of freedom.
Orbital ordering in the perovskite-based mangenites is typically coupled to the lattice through Jahn-Teller distortions of the oxygen octahedra, and often accompanied by Mn$^{3+}$/Mn$^{4+}$ charge ordering as well as antiferromagnetic ordering patterns.
Much of this phenomenology is also encountered among the members of the quadruple perovskite manganites family AMn$_7$O$_{12}$, where A is a 2+ or 3+ cation, which were additionally shown to display interesting dielectric and multiferroic properties~\cite{Vasilev2007,Belik2016}. 

Within this family, CaMn$_{7}$O$_{12}$ has recently attracted a lot of scientific attention due the observation of huge, magnetically-induced ferroelectric polarization (2870 $\mu$Cm$^{-2}$ for a single crystal)~\cite{Zhang2011,Johnson2012}.
Earlier x-ray and neutron scattering studies have revealed that CaMn$_{7}$O$_{12}$ undergoes upon cooling a sequence of phase transitions associated with charge, orbital and spin ordering. 
While at high temperatures CaMn$_{7}$O$_{12}$ crystallizes in a distorted perovskite cubic structure (space group Im$\overline{3}$), at T$_s$$\approx$440 K a first-order structural phase transition takes place towards the rhombohedral R$\overline{3}$ structure (for the followings we will use the hexagonal setting of the R$\overline{3}$ space group)~\cite{Bochu1980,Troyanchuk1997,Vasilev2007}. 
This transition relates to the Jahn-Teller Mn$^{3.25}$ cations of the MnO$_6$ perovskite octahedra (four in each formula unit) and coincides with their simultaneous charge ordering in three Mn$^{3+}$ and one Mn$^{4+}$, occupying non-equivalent sites in the low temperature hexagonal structure.
Upon further cooling, an incommensurate structural distortion with the propagation vector q$_{o}$=(0, 0, 2.077) sets in below T$_o$=250 K, involving a variation of the Mn$^{3+}$-O bond length in the ab-plane~\cite{Przenioslo2004,Slawinski2008}. 
Orbital ordering of the Mn$^{3+}$ cations, following a periodic modulation of the occupation of the 3x$^2$-r$^2$ and the 3y$^2$-r$^2$ orbitals, was inferred from the modulated Mn$^{3+}$-O bond lengths~\cite{Perks2012}.
 
Ferroelectricity appears below T$_{N1}$=90 K, simultaneously with the antiferromagnetic ordering of the manganese magnetic moments. 
The modulation wavevector of the observed helical magnetic pattern locks-in to the one of the preceding orbital-ordering, q$_{m}$=q$_{o}$/2, resulting in an intertwined magneto-orbital helix~\cite{Slawinski2010,Johnson2012}.
The observed magnetic structure was suggested to arise from the effect of the orbital modulation on the magnetic exchange interactions~\cite{Perks2012}.
The simultaneous apparition of a huge electric polarization perpendicular to the spin rotation plane of the helical structure has been attributed to the coupling between the magneto-orbital helix and the global structural rotation~\cite{Zhang2011,Johnson2012}.
However, very recently doubts were reported concerning the appearance of an intrinsic ferroelectric polarization in this compound~\cite{Noriki2016}.
Finally, below T$_{N2}$=48 K another magnetic transition takes place, described by multiple distinct magnetic propagation vectors~\cite{Slawinski2012,Johnson2016} 

\begin{figure*}
\includegraphics[width=0.98\textwidth]{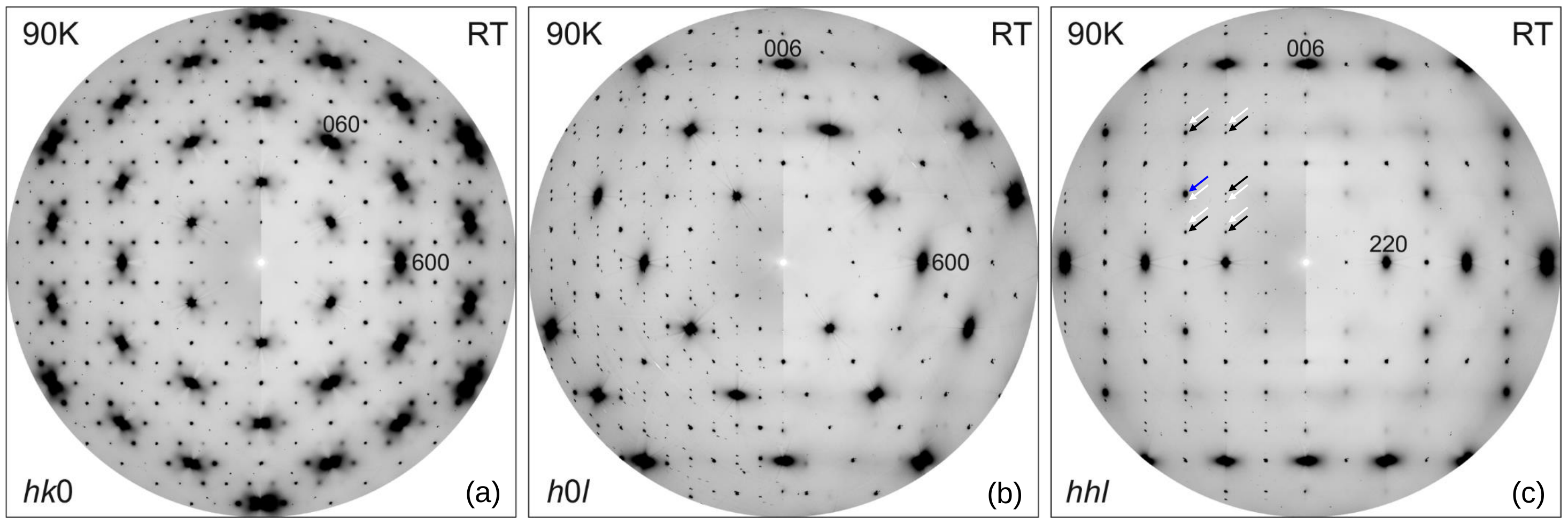}
\caption{Reconstructed reciprocal space maps of the (a) (\textit{hk}0), (b) (\textit{h}0\textit{l}) and (c) (\textit{hhl}) planes from DS data taken at room temperature (right half of panels) and 90 K (left half of the panels). \textit{h}, \textit{k} and \textit{l} are given in reciprocal lattice units in the hexagonal setting. 
Black (white) arrows in panel (c) indicate the first (second) order CDW satellite peaks. The blue arrow indicates the superstructure position investigated with IXS. In panel (a) the diffuse peaks originate from tails of out of plane satellite peaks. In panel (b) the extra spots not-lying along the \textit{h}0\textit{l} lines with integer \textit{h} are due to a minority twin.} \label{Fig1}
\end{figure*}

Motivated by the importance of the structural modulation and the associated orbital order for the stabilization of the magneto-orbital helix and the emergence of magnetically-induced ferroelectricity, we investigated the lattice dynamics across the transition at T$_o$. 
Earlier works have studied the phononic response to the low temperature orbital and magnetic ordering by means of Raman spectroscopy~\cite{Du2014,Iliev2014,Nonato2014}. 
Below T$_o$, new phonon modes appear in the Raman spectra, originating from the folding of the Brillouin zone in the incommensurately modulated structure~\cite{Du2014,Iliev2014}.
In addition, Du \textit{et al.} have identified a low temperature soft vibrational mode, emerging from zero energy at T$_o$ and have associated it to the amplitude excitations of the order parameter, suggesting the soft phonon nature of the structural modulation.
Here we present the first momentum-resolved single crystal lattice dynamical investigation of  CaMn$_{7}$O$_{12}$, through a combination of thermal diffuse scattering (DS) and high energy resolution inelastic x-ray scattering (IXS) measurements. 
In our data we clearly observe intense diffuse scattering features and temperature-dependent phonon softening well above T$_{o}$, therefore unambiguously establishing the soft-mode driven nature of the transition at T$_o$. 
The momentum width of the phonon anomaly is compatible with both momentum-dependent electron-phonon coupling as well as competing interactions (or "frustration") as the driving force for the incommensurate modulation.

\section{\label{sec:Experimental Methods}II. Experimental Methods}
Single crystals of CaMn$_{7}$O$_{12}$ were grown by a flux method as described elsewhere~\cite{Johnson2012}. The high crystal quality was confirmed by x-ray diffraction and is visible in the reconstructed reciprocal space maps of Fig.~\ref{Fig1}. The lattice parameters in the hexagonal setting are \textit{a$_{h}$} = \textit{b$_{h}$} = 10.4577 \r{A} and \textit{c$_{h}$} = 6.3422 \r{A}. The crystals were mechanically detwinned by uniaxial compression at room temperature as described in reference~\cite{Yuan2015}. The samples were characterized further by transport measurements in an earlier work~\cite{Du2014}. In order to access a large portion of the reciprocal space the experiments were performed in the transmission geometry using samples with thicknesses close to the x-ray absorption length ($\sim$300 $\mu$m at x-ray energies of $\sim$16-18 keV). 
 
X-ray thermal DS measurements were taken at the ID23 beamline of the European Synchrotron Radiation Facility (ESRF), using a monochromatic x-ray beam of 17.7 keV. DS data were recorded with a Pilatus 6M detector, at room temperature and at 90 K using a cryostream cooling system. The CrysAlis software package was used to obtain the experimental orientation matrix and to perform a preliminary data evaluation~\cite{Crysalis}. The reconstruction of selected reciprocal space layers was prepared using a locally developed software. Symmetry operations of the $\overline{3}$m point group were applied to the scattering patterns in order to improve the signal-to-noise ratio and to remove the gaps of the detector elements. 

IXS measurements were performed at the ID28 beamline of the ESRF with an incident photon energy of 17.794 keV and a corresponding instrumental energy resolution of 3 meV. The x-ray beam was focused by multilayer mirrors to a spot of 50 $\times$ 30 $\mu$m (horizontally $\times$ vertically) on a CaMn$_{7}$O$_{12}$ single crystal. The momentum transfer was selected by the scattering angle and the sample orientation and the momentum resolution was set to $\sim$0.25 nm$^{-1}$ in the scattering plane and 0.75 nm$^{-1}$ perpendicular to it. Energy scans were recorded from -20 to 20 meV, at selected temperatures from 275 to 490 K, using an Oxford Cryostream 700 Plus cooling and heating system. The experimental data were fitted using standard damped harmonic oscillator functions convoluted with the experimental resolution~\cite{Fak1997,Krisch2007}.
%to 0.043 r.l.u. in the a*(b*) directions and 0.026 r.l.u. along the c* direction of the reciprocal space. 

\begin{figure*}
\includegraphics[width=0.98\textwidth]{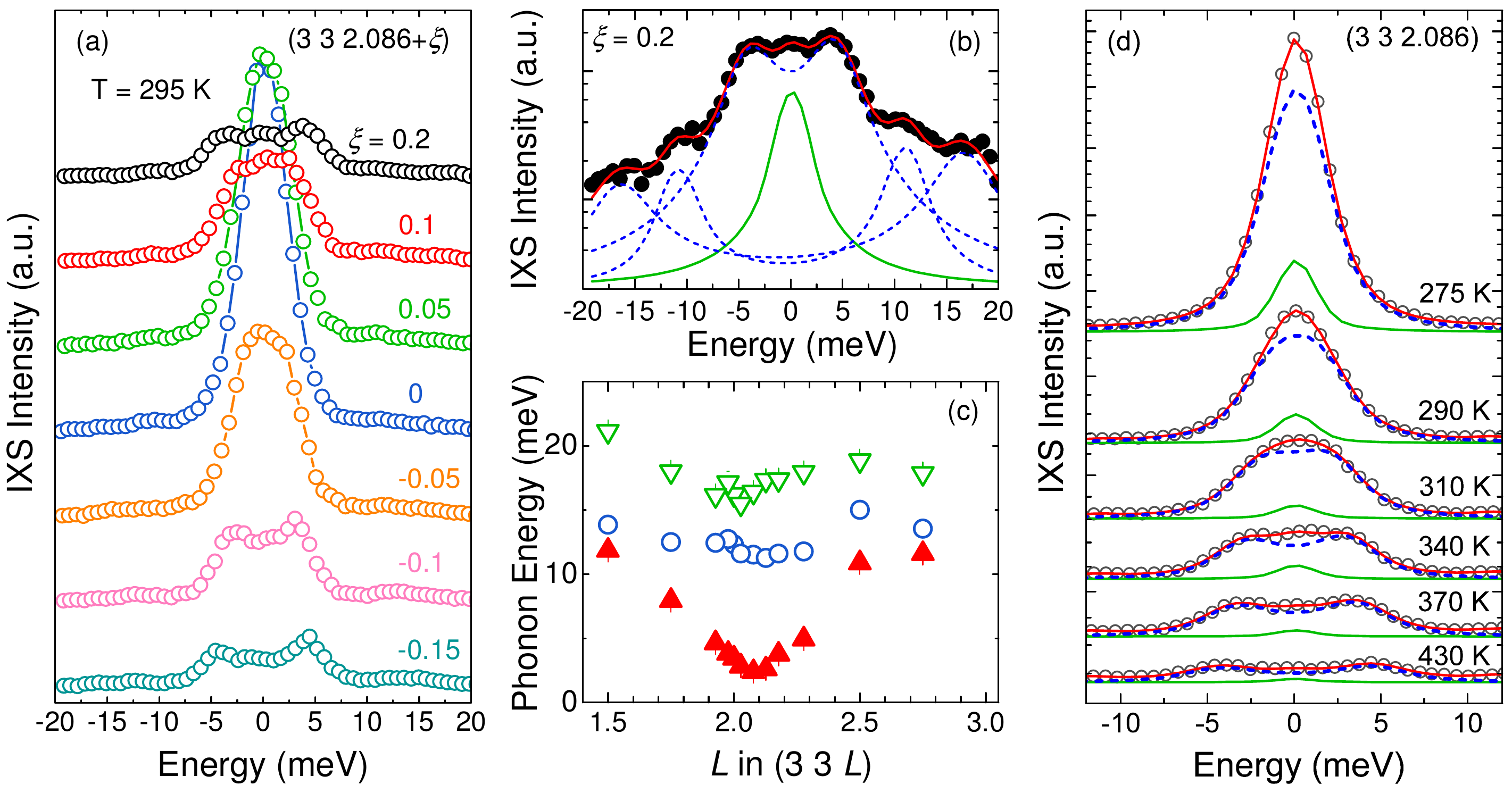}
\caption{(a) Room temperature IXS spectra at (\textit{h k l}) = (3 3 2.086+\textit{$\xi$}). The scans are vertically shifted for clarity. The statistical error bars are smaller than the data symbols. (b) IXS spectrum at the reciprocal space point (3 3 2.286) (\textit{$\xi$}=0.2). The dashed blue and solid green lines show the results of a fit of the inelastic and elastic part of the spectrum respectively, while the red line represents the total fit result. For clarity reasons, the vertical scale is logarithmic. (c) Phonon energy dispersions of the three phonons of (a,b) at room temperature. (d) Temperature dependence of the raw IXS spectra taken at q$_{o}$ = (3 3 2.086). The y-axis is in a logarithmic scale and the scans are vertically shifted for clarity. The statistical error bars are smaller than the data symbols. The dotted green (blue) lines represent the results of fits to the elastic line (low energy phonon) intensity, and the solid red lines represent the total fit.} \label{Fig2}
\end{figure*}

\section{\label{sec:Experimental Results}III. Experimental Results}
In order to identify favorable reciprocal space positions for the study of the charge ordered state, we started our investigation with a DS study, similar to what was recently done in the case of transition metal dichalcogenides and cuprate superconductors~\cite{Leroux2012,LeTacon2013,Bosak2015}.
Fig.~\ref{Fig1} shows reconstructed reciprocal space maps of selected planes in the non-modulated (room temperature - right panels) and the modulated (90 K - left panels) structures. 

The incommensurate structural distortion is clearly reflected in the DS data: at 90 K, in addition to the crystallographically expected reflections of the R$\overline{3}$ space group (reflection conditions: -\textit{h}+\textit{k}+\textit{l}=3\textit{n}), new diffraction spots appear, as shown for the (\textit{h}0\textit{l}) and (\textit{hhl}) planes in Fig.~\ref{Fig1}-b,c. 
In agreement with previous x-ray studies~\cite{Przenioslo2004,Slawinski2008,Slawinski2009}, the new features run along the c* direction at the positions (\textit{h k l}+q$_{o}$), where q$_{o}$=2.086. 
We note that up to second order incommensurate superstructure reflections are clearly observed in the low temperature DS data (see black and white arrows in Fig.~\ref{Fig1}-c). 
At 90 K, the superstructure peaks are sharply localized in momentum space, indicating long correlation lengths of the charge order modulation below T$_{o}$.    
On the other hand, at room temperature, and therefore already well above T$_{o}$, a series of intense albeit broad (0.2\textit{c$^{*}_{h}$} $\times$ 0.12(\textit{a$^{*}_{h}$}+\textit{b$^{*}_{h}$}) = 1.98 $\times$ 1.48 nm$^{-1}$) DS features are observed at the positions of the low temperature superstructure peaks.  

To get further insights about these diffuse features we performed temperature dependent IXS measurements across their positions. 
For the followings, we focus on the q$_{o}$=(3 3 2.086) DS peak which shows the highest intensity within the experimentally accessible range of the reciprocal space (blue arrow in Fig.~\ref{Fig1}-c). 
The room temperature dispersion of the raw IXS spectra across this peak and along the modulation vector is shown in Fig.~\ref{Fig2}-a. 
Within the studied energy range, the IXS spectra are composed of a central peak at zero energy loss and three optical phonon peaks in the Stokes and anti-Stokes parts of the spectra, with their intensity ratios following accurately the detailed balance (see Fig.~\ref{Fig2}-b for the details of the fitting). 
The linewidths of the observed peaks are not resolution limited, therefore we cannot exclude that each peak consists of more than one unresolved phonons. 
Focusing on the lowest energy optical phonon branch, a pronounced softening from $\sim$10 to $\sim$2.5 meV is observed upon approaching q$_{o}$ from the (3 3 3) Brillouin zone center (see Fig.~\ref{Fig2}-c for the full dispersion of the phonon energies). 
As will be discussed in detail further on, the q-width of the the soft phonon anomaly is in good agreement with the room temperature DS feature width at q$_{o}$ (Fig.~\ref{Fig1}-c). Moreover the q-cut shape of the DS feature is close to a lorentzian, which corresponds to a nearly conical phonon dispersion as seen experimentally in the IXS data.  
Within our experimental energy resolution we did not observe any clear softening in the higher energy phonon modes.

The temperature dependence of the soft phonon upon approaching T$_o$ from above is traced in Fig.~\ref{Fig2}-d.
In contrast to the expected anharmonic phonon hardening at low temperatures, the low energy optical phonon at q$_{o}$ continuously softens from $\sim$5 meV at 430 K to less than $\sim$1.75 meV at 275 K, appearing at this temperature as a broad quasielastic peak merged with the elastic line (within the experimental energy resolution). 
Below 275 K the quasielastic peak dominates the IXS spectra and below T$_o$ the strong superstructure peak of the modulated structure prevents any low energy phonon observation. 
The temperature dependence of the extracted low energy phonon dispersion is summarized in Fig.~\ref{Fig3}. 
The dip at q$_{o}$ is present throughout the whole studied temperature range, starting from a $\sim$1.8 meV softening at 400 K, and therefore 150 K above the transition temperature T$_o$, and increasing up to $\sim$3.5 meV at 275 K. 

We note at this point that for temperatures higher than 440 K the IXS phonon spectrum changes drastically (see Fig.~\ref{Fig4} for T=460 K), signaling the structural transition to the high temperature cubic Im$\overline{3}$ phase. Similar changes in the phononic spectra above 440 K have been reported in earlier Raman, infrared and THz spectroscopic studies~\cite{Iliev2014,Kadlec2014}. Our IXS data in the temperature range 440-490 K show no soft phonon behavior in the high temperature phase, in agreement with the previously reported first order character of this Jahn-Teller driven structural phase transition~\cite{Bochu1980,Przenioslo2002,Vasilev2007}.   

\section{\label{sec:Discussion}IV. Discussion}

\begin{figure}
\includegraphics[width=0.9\linewidth]{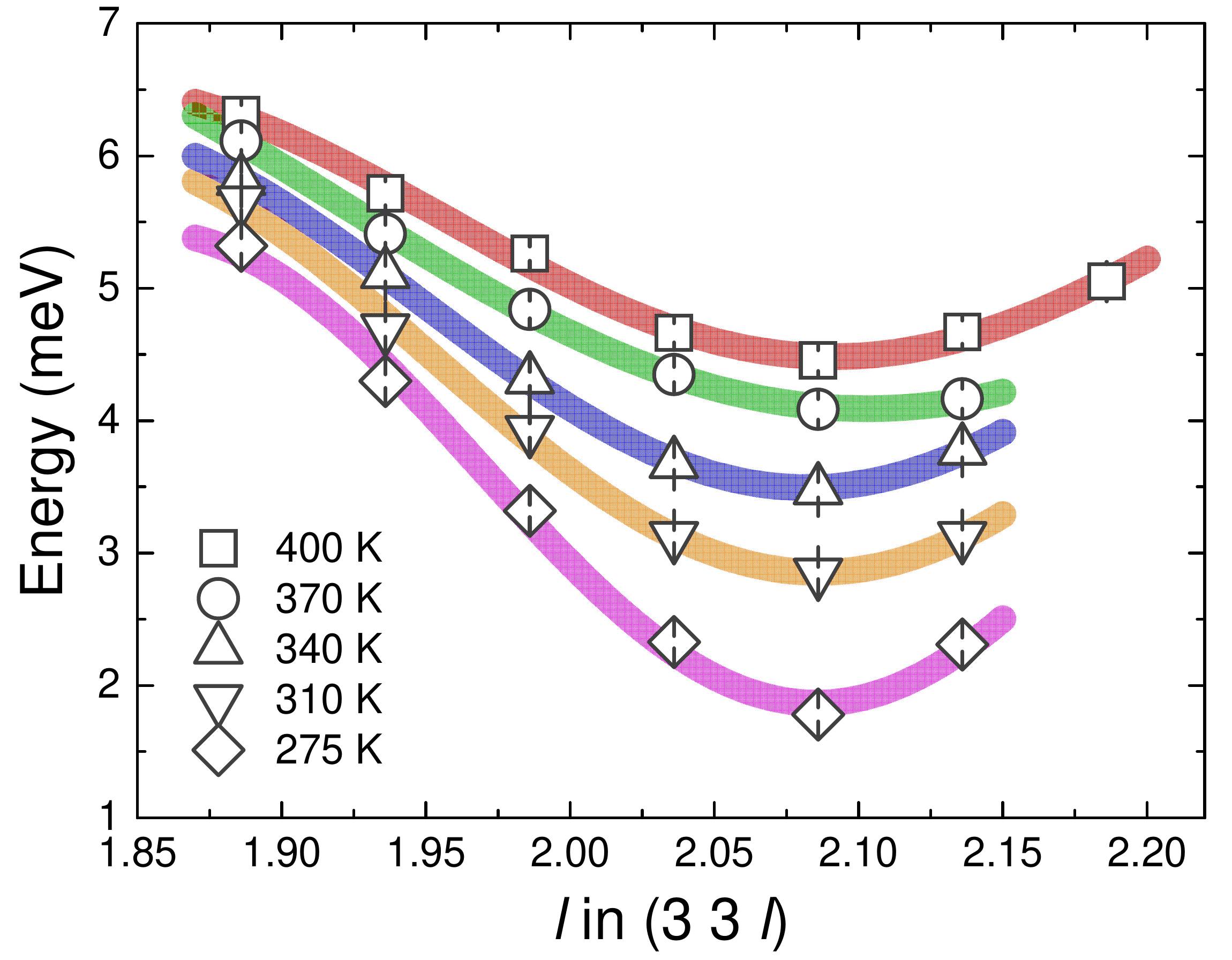}
\caption{Momentum and temperature dependence of the phonon energy at T= 400, 370, 340, 310 and 275 K. The solid lines correspond to a sinusoidal model description as a guide to the eye.} \label{Fig3}
\end{figure}

In the canonical description of soft-phonon driven displacive phase transitions, the energy of a phonon of the high temperature and high symmetry phase continuously decreases (softens) upon cooling towards the transition temperature T$_o$~\cite{Scott1974,Dove1997}. The temperature dependence of the phonon energy (frequency) follows:
\begin{equation}\label{eq1}
\omega=\omega_o(1-T/T_o)^{\gamma}
\end{equation}
with $\gamma$=0.5 within the mean-field picture. At T$_o$ the phonon energy reaches zero and therefore the restoring force against the phononic distortion vanishes and the atomic displacements associated with the soft phonon mode freeze in the structure of the low temperature and low symmetry phase.
The temperature dependence of the soft phonon mode in CaMn$_{7}$O$_{12}$ is summarized in Fig.~\ref{Fig5}. 
The IXS results are fitted to Eq.~\ref{eq1}, with $\omega_o$=47.03$\pm$0.65 cm$^{-1}$, T$_o$=250.0$\pm$5.5 K and $\gamma$=0.5$\pm$0.04, following thus very accurately the canonical mean-field power law. 

\begin{figure}
\includegraphics[width=0.7\linewidth]{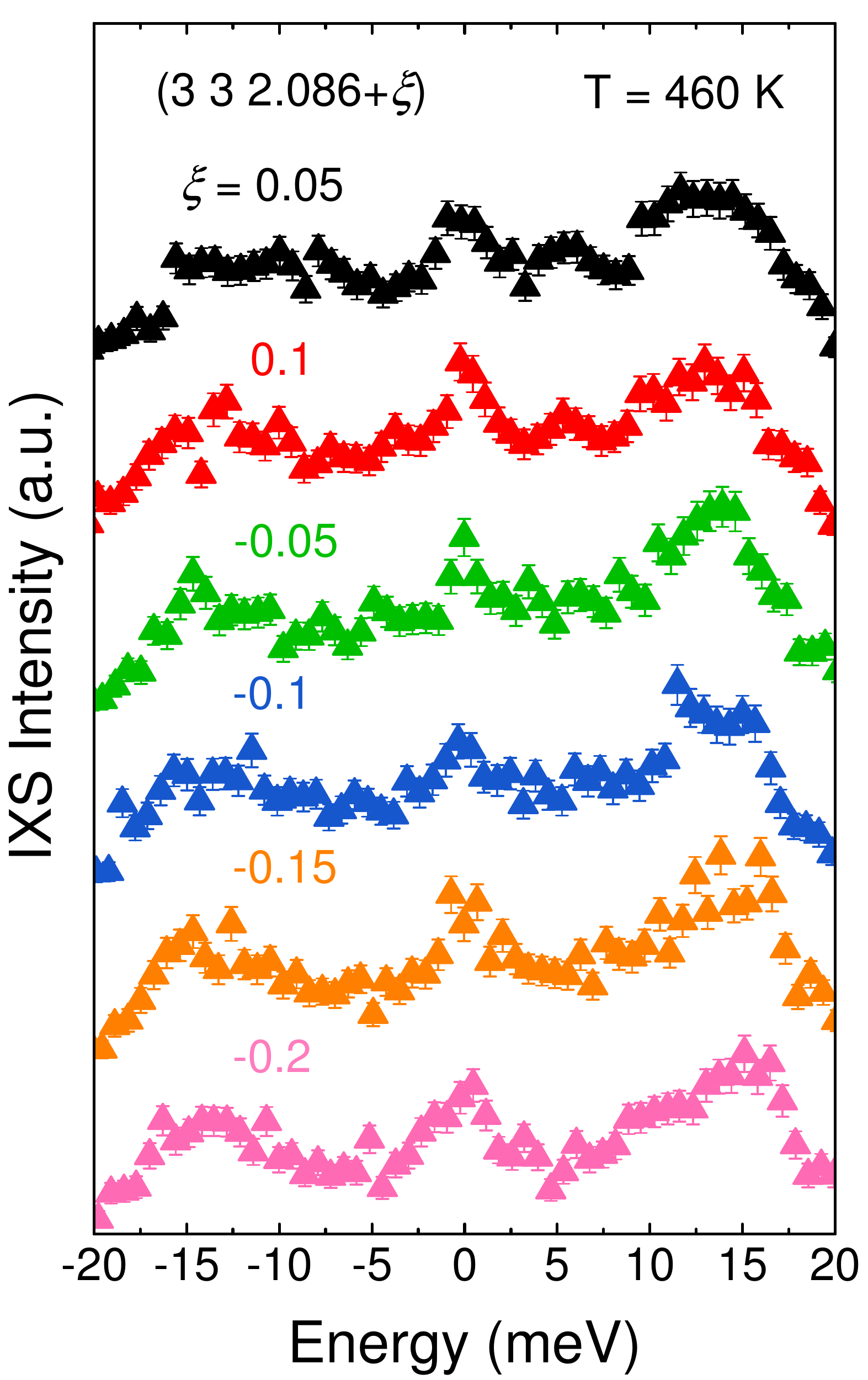}
\caption{IXS spectra at (h k l) = (3 3 2.086+$\xi$) recorded at 460 K. The scans are vertically shifted for clarity.} \label{Fig4}
\end{figure}
 
In addition to extra phonon peaks in the Raman spectra due to the back-folding of the phonon dispersions in the incommensurately distorted structure, Du \textit{et al.} reported the appearance of a new soft vibrational mode below T$_o$ and associated it to the amplitude excitation of the composite order parameter~\cite{Du2014,Cummins1990}. 
The new mode continuously softens as T$_o$ is approached from below, following a canonical power-law temperature dependence with  $\omega_o$=81.2 cm$^{-1}$ and T$_o$=249.4 K (blue symbols and line in Fig.~\ref{Fig5}), and disappears above T$_o$. 
While the finite momentum of the orbital order does not allow the observation of the soft phonon mode in the Raman spectra above T$_o$, our momentum resolved results offer direct evidence that the transition at T$_o$ is due to a soft phonon lattice instability.
To the best of our knowledge this is the first experimental evidence for a soft phonon driven orbital order in a magnanite.

\begin{figure}
\includegraphics[width=0.9\linewidth]{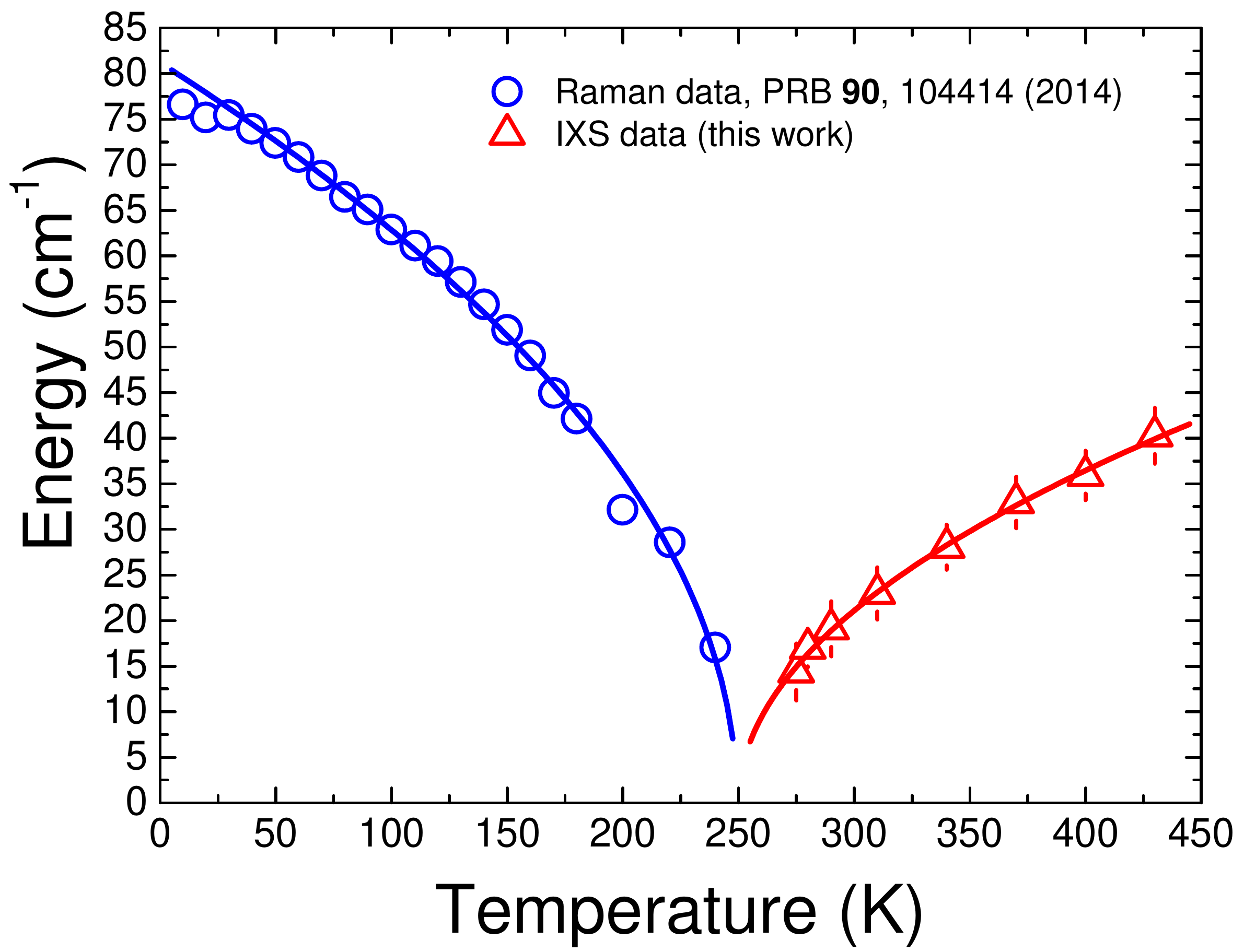}
\caption{Temperature dependence of the soft-mode energy. The blue symbols are taken from the Raman scattering data of Du \textit{et al.}~\cite{Du2014} below T$_{o}$ and the red symbols are the results of the current IXS study above T$_{o}$. The solid lines represent power low fits to the experimental data (see discussion in main text).} \label{Fig5}
\end{figure}
 
In metallic systems, incommensurate charge density waves (CDW) are usually understood as "nesting"-driven Fermi surface instabilities, with the nesting wavevector selecting the CDW wavevector q$_{CDW}$~\cite{Gruner1988}, which in the presence of electron-phonon coupling lead to Kohn anomalies in the vibrational spectra~\cite{Kohn1959}.
Alternatively, it has been argued that strongly q-dependent electron-phonon coupling, instead of Fermi surface nesting, could be the driving force for the incommensurate charge ordering and the one dictating the ordering wavevector~\cite{Johannes2008,Zhu2015}. 
For the well studied classic CDW system NbSe$_{2}$, it has been claimed that the experimentally observed phonon anomalies allow to distinguish between the two possible CDW mechanisms, since the extended q-width of the phonon renormalizations is not in line with the sharply localized dips expected in a Fermi surface nesting picture~\cite{Weber2011b}.

Interestingly, the phonon softening in the IXS data of CaMn$_{7}$O$_{12}$ appears to be relatively broad in momentum space, with a q-width (at half-depth of the conical dispersion) of $\sim$0.4 r.l.u. along the \textit{c$^{*}_{h}$} direction, and therefore $\sim$13\% of the Brillouin zone extent along this direction (with  \textit{c$_{h}$}=\textit{a$_{r}$}+\textit{b$_{r}$}+\textit{c$_{r}$}, where \textit{a$_{r}$}, \textit{b$_{r}$} and \textit{c$_{r}$} are the lattice parameters in the rhombohedral setting). A significant softening amplitude (18\% of the phonon energy) is observed even 0.2 r.l.u. away from q$_{o}$ (see Fig.~\ref{Fig2} and Fig.~\ref{Fig3}).
In the case of insulating systems, such as CaMn$_{7}$O$_{12}$, incommensurate lattice distortions are typically considered as a result of competing interactions, like long range Coulomb interactions or short range exchange interactions which favor different periodicities~\cite{Cummins1990,Monceau2012}, or some form of geometric frustration~\cite{Khomskii2014}. From the lattice dynamics point of view, in the classic ionic insulator K$_{2}$SeO$_{4}$, phonon anomalies centered around the incommensurate wavevector were found to span over the bigger part of the Brillouin zone~\cite{Iizumi1977}. 
Nevertheless, microscopic mechanisms involving strongly momentum-dependent electron-phonon coupling, similar to the CDW mechanism described above for metallic systems, may also be relevant for insulators and would produce phonon anomalies with a momentum width tightly linked to the wavevector dependence of the electron-phonon coupling~\cite{Du2014}.
Therefore, whereas a competing interactions approach would be relevant in this Jahn-Teller distorted system, a q-dependent electron-phonon coupling could also account for the observed momentum range of the softening in CaMn$_{7}$O$_{12}$, or contributions from both mechanisms should be taken into consideration.
 
Regarding the nature of the soft phonon mode, in the absence of non-magnetic lattice dynamical calculations we cannot associate it to a specific atomic displacement pattern.
Nevertheless, given that the transition is of second order and soft phonon driven, it is natural to associate the frozen phonon displacement pattern to the low temperature modulated structure. 
Moreover, a comparison of the IXS spectra with the zone center phonons shows that, while in the R$\overline{3}$ structure there are no Raman active modes below $\sim$22 meV~\cite{Iliev2014}, the observed IXS phonon energies match well with the reported infrared vibrations at $\sim$11.4, 12.8 and 18 meV~\cite{Kadlec2014}.

The same infrared study combined with neutron scattering data, reports the possible coupling of polar phonon and spin modes ("electromagnons") below the magnetic ordering temperatures T$_{N1}$ and T$_{N2}$~\cite{Kadlec2014}. Anomalies in the temperature dependence of the frequencies and linewidths have also been reported for the Raman active phonon modes upon entering the magnetically ordered states. While these could be related to anomalies in the lattice parameters across the magnetic transition at T$_{N2}$~\cite{Sanchez2009,Przenioslo2004}, in the absence of structural anomalies across T$_{N1}$ the phonon anomalies across T$_{N1}$ were attributed to direct spin-phonon coupling~\cite{Nonato2014}.  
Moreover, a recent density functional theory study based on the experimental magnetic structure showed the existence of unstable modes in a large part of the Brillouin zone and indicated the major contribution of A$_g$ symmetry Raman type lattice distortions in the low temperature ferroelectric polarization~\cite{Dai2015}.
While the current IXS study has been limited to temperatures above T$_{o}$, a future momentum resolved phononic study across the magnetic transitions would provide further valuable insights regarding the lattice contribution in their microscopic mechanism.  

\section{\label{sec:Conclusions}V. Conclusions}

In summary, we investigated the phase transition mechanism of the incommensurate charge and orbital ordering in CaMn$_{7}$O$_{12}$, through a detailed temperature dependent study of the lattice dynamics by means of momentum resolved inelastic x-ray scattering.
In our preparatory diffuse scattering measurements we observe intense diffuse features at the superstructure positions of the incommensurate structure, already well above the orbital ordering temperature T$_o$.
Inelastic x-ray scattering scans across these features reveal a pronounced optical phonon softening centered around the superstructure positions.
The softening increases continuously as T$_o$ is approached from above following a canonical power law, and therefore unambiguously establishing that the phase transition is of second order and soft-phonon driven. 
The observation of the phonon anomaly over a broad momentum regime appears compatible with microscopic mechanisms of the incommensurate modulation deriving from q-dependent electron-phonon coupling and also from competing interactions in this Jahn-Teller distorted insulator. 
Our data demonstrate the pivotal role played by the lattice in mediating the orbital order and therefore the realization of the tightly coupled magnetic order and magnetically-driven ferroelecticity in this system and call for further phonon investigations in orbitally ordered magnanites.

\section{Acknowledgement}
We thank the Macromolecular Crystallography (ID23) Beamline at ESRF for beamtime allocation for the diffuse scattering data.

%%%%%%%%%%%%%%%%%%%%%%BIBLIOGRAPHY%%%%%%%%%%%%%%%%%%%%%%%%%%%%%%%%%%%%%%%%%%%%%%%


\begin{thebibliography}{}

\bibitem{Vasilev2007}
A.~N. Vasil'ev and O.~S. Volkova, \href{http://dx.doi.org/10.1063/1.2747047} {Low Temp. Phys. $\textbf{33}$, 895~(2007).}

\bibitem{Belik2016}
A.~A. Belik, Y.~S. Glazkova, Y. Katsuya, M. Tanaka, A.~V. Sobolev and I.~A. Presniakov, \href{http://dx.doi.org/10.1021/acs.jpcc.6b01649} {J. Phys. Chem. C $\textbf{120}$, 8278~(2016).}

\bibitem{Zhang2011}
G. Zhang, S. Dong, Z. Yan, Y. Guo, Q. Zhang, S. Yunoki, E. Dagotto and J.-M. Liu, \href{http://dx.doi.org/10.1103/PhysRevB.84.174413} {Phys. Rev. B $\textbf{84}$, 174413~(2011).}

\bibitem{Johnson2012}
R.~D. Johnson, L.~C. Chapon, D.~D. Khalyavin, P. Manuel, P.~G. Radaelli and C. Martin, \href{http://dx.doi.org/10.1103/PhysRevLett.108.067201} {Phys. Rev. Lett. $\textbf{108}$, 067201~(2012).}

\bibitem{Bochu1980}
B. Bochu, J. Buevoz, J. Chenavas, A. Collomb, J. Joubert and M. Marezio, \href{http://dx.doi.org/10.1016/0038-1098(80)90668-7} {Solid State Commun. $\textbf{36}$, 133~(1980).}

\bibitem{Troyanchuk1997}
I. Troyanchuk and A. Chobot, Crystallogr. Rep. $\textbf{42}$, 983~(1997).

\bibitem{Przenioslo2004}
R. Przenios{\l}o, I. Sosnowska, E. Suard, A. Hewat and A. Fitch, \href{http://dx.doi.org/10.1016/j.physb.2003.10.013} {Physica B $\textbf{344}$, 358~(2004).}

\bibitem{Slawinski2008}
W. S{\l}awi{\'{n}}ski, R. Przenios{\l}o, I. Sosnowska, M. Bieringer, I. Margiolaki, A.~N. Fitch and E. Suard, \href{http://dx.doi.org/10.1088/0953-8984/20/10/104239} {J. Phys.: Condens. Matter $\textbf{20}$, 104239~(2008).}

\bibitem{Perks2012}
N. Perks, R. Johnson, C. Martin, L. Chapon and P. Radaelli, \href{http://dx.doi.org/10.1038/ncomms2294} {Nat. Commun. $\textbf{3}$, 1277~(2012).}

\bibitem{Slawinski2010}
W. S{\l}awi{\'{n}}ski, R. Przenios{\l}o, I. Sosnowska and M. Bieringer, \href{http://dx.doi.org/10.1088/0953-8984/22/18/186001} {J. Phys.: Condens. Matter $\textbf{22}$, 186001~(2010).}

\bibitem{Noriki2016}
N. Terada, Y.~S. Glazkova and A.~A. Belik, \href{http://dx.doi.org/10.1103/PhysRevB.93.155127} {Phys. Rev. B $\textbf{93}$, 155127~(2016).}

\bibitem{Slawinski2012}
W. S{\l}awi{\'{n}}ski, R. Przenios{\l}o, I. Sosnowska and A. Chrobak, \href{http://dx.doi.org/10.1143/JPSJ.81.094708} {J. Phys. Soc. Jpn $\textbf{81}$, 094708~(2012).}

\bibitem{Johnson2016}
R.~D. Johnson, D.~D. Khalyavin, P. Manuel, A. Bombardi, C. Martin, L.~C. Chapon and P.~G. Radaelli, \href{http://dx.doi.org/10.1103/PhysRevB.93.180403} {Phys. Rev. B $\textbf{93}$, 180403~(2016).}

\bibitem{Du2014}
X. Du, R. Yuan, L. Duan, C. Wang, Y. Hu and Y. Li, \href{http://dx.doi.org/10.1103/PhysRevB.90.104414} {Phys. Rev. B $\textbf{90}$, 104414~(2014).}

\bibitem{Iliev2014}
M.~N. Iliev, V.~G. Hadjiev, M.~M. Gospodinov, R.~P. Nikolova and M.~V. Abrashev, \href{http://dx.doi.org/10.1103/PhysRevB.89.214302} {Phys. Rev. B $\textbf{89}$, 214302~(2014).}

\bibitem{Nonato2014}
A. Nonato, B.~S. Araujo, A.~P. Ayala, A.~P. Maciel, S. Yanez-Vilar, M. Sanchez-Andujar, M.~A. Senaris-Rodriguez and C.~W.~A. Paschoal, \href{http://dx.doi.org/10.1063/1.4902234} {Appl. Phys. Lett. $\textbf{105}$, 222902~(2014).}

\bibitem{Yuan2015}
R. Yuan, L. Duan, X. Du and Y. Li, \href{http://dx.doi.org/10.1103/PhysRevB.91.054102} {Phys. Rev. B $\textbf{91}$, 054102~(2015).}

\bibitem{Crysalis}
$\textit{CrysAlis PRO}$ (Rigaku Oxford Diffraction, Yarnton, England, 2015)


\bibitem{Fak1997}
B. F\r{a}k and B. Dorner, \href{http://dx.doi.org/10.1016/S0921-4526(97)00121-X} {Physica B $\textbf{234}$, 1107~(1997).}

\bibitem{Krisch2007}
M. Krisch and F. Sette, {Light Scattering in Solids IX} $\textbf{108}$, 317~(2007).

\bibitem{Leroux2012}
M. Leroux, M. Le~Tacon, M. Calandra, L. Cario, M.-A. M\'easson, P. Diener, E. Borrissenko, A. Bosak and P. Rodi\`ere, \href{http://dx.doi.org/10.1103/PhysRevB.86.155125} {Phys. Rev. B $\textbf{86}$, 155125~(2012).}

\bibitem{LeTacon2013}
M. Le~Tacon, A. Bosak, S.~M. Souliou, G. Dellea, T. Loew, R. Heid, K.-P. Bohnen, G. Ghiringhelli, M. Krisch and B. Keimer, \href{http://dx.doi.org/10.1038/NPHYS2805} {Nat. Phys. $\textbf{10}$, 52~(2014).}

\bibitem{Bosak2015}
A. Bosak, D. Chernyshov, B. Wehinger, B. Winkler, M. Le~Tacon and M. Krisch, \href{http://dx.doi.org/10.1088/0022-3727/48/50/504003} {J. Phys. D: Appl. Phys. $\textbf{48}$, 504003~(2015).}

\bibitem{Slawinski2009}
W. S{\l}awi{\'{n}}ski, R. Przenios{\l}o, I. Sosnowska, M. Bieringer, I. Margiolaki and E. Suard, \href{http://dx.doi.org/10.1107/S0108768109025300} {Acta Cryst. $\textbf{65}$, 535~(2009).}

\bibitem{Kadlec2014}
F. Kadlec, V. Goian, C. Kadlec, M. Kempa, P. Van\v{e}k, J. Taylor, S. Rols, J. Prokle\v{s}ka, M. Orlita and S. Kamba, \href{http://dx.doi.org/10.1103/PhysRevB.90.054307} {Phys. Rev. B $\textbf{90}$, 054307~(2014).}

\bibitem{Przenioslo2002}
R. Przenios{\l}o, I. Sosnowska, E. Suard, A. Hewat and A.~N. Fitch, \href{http://dx.doi.org/10.1088/0953-8984/14/23/308} {J. Phys.: Condens. Matter $\textbf{14}$, 5747~(2002).}

\bibitem{Scott1974}
J.~F. Scott, \href{http://dx.doi.org/10.1103/RevModPhys.46.83} {Rev. Mod. Phys. $\textbf{46}$, 83~(1974).}

\bibitem{Dove1997}
M.~T. Dove, \href{http://dx.doi.org/10.2138/am-1997-3-401} {Am. Mineral. $\textbf{82}$, 213~(1997).}

\bibitem{Cummins1990}
H.~Z. Cummins, \href{http://dx.doi.org/10.1016/0370-1573(90)90058-A} {Phys. Rep. $\textbf{185}$, 211~(1990).}

\bibitem{Gruner1988}
G. Gr\"uner, \href{http://dx.doi.org/10.1103/RevModPhys.60.1129} {Rev. Mod. Phys. $\textbf{60}$, 1129~(1988).}

\bibitem{Kohn1959}
W. Kohn, \href{http://dx.doi.org/10.1103/PhysRevLett.2.393} {Phys. Rev. Lett. $\textbf{2}$, 393~(1959).}

\bibitem{Johannes2008}
M.~D. Johannes and I.~I. Mazin, \href{http://dx.doi.org/10.1103/PhysRevB.77.165135} {Phys. Rev. B $\textbf{77}$, 165135~(2008).}

\bibitem{Zhu2015}
X. Zhu, Y. Cao, J. Zhang, E.~W. Plummer and J. Guo, \href{http://dx.doi.org/10.1073/pnas.1424791112} {Proc. Natl. Acad. Sci. $\textbf{112}$, 2367~(2015).}

\bibitem{Weber2011b}
F. Weber, S. Rosenkranz, J.-P. Castellan, R. Osborn, R. Hott, R. Heid, K.-P. Bohnen, T. Egami, A.~H. Said and D. Reznik, \href{http://dx.doi.org/10.1103/PhysRevLett.107.107403} {Phys. Rev. Lett. $\textbf{107}$, 107403~(2011).}

\bibitem{Monceau2012}
P. Monceau, \href{http://dx.doi.org/10.1080/00018732.2012.719674} {Adv. Phys. $\textbf{61}$, 325~(2012).}

\bibitem{Khomskii2014}
D.~I. Khomskii, \textit{Transition Metal Compounds} (Cambridge University Press~2014).

\bibitem{Iizumi1977}
M. Iizumi, J.~D. Axe, G. Shirane and K. Shimaoka, \href{http://dx.doi.org/10.1103/PhysRevB.15.4392} {Phys. Rev. B $\textbf{15}$, 4392~(1977).}

\bibitem{Sanchez2009}
M. M.~S{\'a}nchez-And{\'{u}}jar, S. Y{\'{a}}{\~{n}}ez-Vilar, N. Biskup, S. Castro-Garc\'{\i}a, J. Mira, J. Rivas and M. Se{\~{n}}ar\'{\i}s-Rodr\'{\i}guez, \href{http://dx.doi.org/10.1016/j.jmmm.2009.02.018} {J. Magn. Magn. Mater. $\textbf{321}$, 1739~(2009).}

\bibitem{Dai2015}
J.-Q. Dai, Y.-M. Song and H. Zhang, \href{http://dx.doi.org/10.1088/1367-2630/17/11/113038 } {New J. Phys. $\textbf{17}$, 113038~(2015).}

\end{thebibliography}
\end{document}